\documentclass[aps]{JHEP3}

\usepackage{graphicx}
\usepackage{dcolumn}
\usepackage{amsfonts}
\usepackage{amsmath}

\newcommand{\ep}{\epsilon}

\newcommand{\td}{\tilde}

\newcommand{\beq}[1]{\begin{eqnarray}\label{#1}}
\newcommand{\eeq}{\end{eqnarray}}

\newcommand{\gym}{g_{_{\rm YM}}}

\newcommand{\CN}[1]{{\cal N}=#1}

\title{Integrable Open Spin Chain in Super Yang-Mills
and the Plane-wave/SYM Duality}
\author{Bin Chen$^1$, Xiao-Jun Wang$^2$ and Yong-Shi Wu$^3$ \\
$^1$ Interdisciplinary Center for Theoretical Study \\ Chinese
Academy of Sciences, Beijing 100080,
China \\
$^2$ Interdisciplinary Center for Theoretical Study \\
University of Science and Technology of China \\
An-Hui, He-Fei 230026, China \\
$^3$ Department of Physics, University of Utah \\
Salt Lake City, Utah 84112, USA \\
E-mail: bchen@itp.ac.cn, wangxj@ustc.edu.cn, wu@physics.utah.edu}

\abstract{We investigate the integrable structures in an $\CN{2}$
superconformal $Sp(N)$ Yang-Mills theory with matter, which is
dual to an open+closed string system. We restrict ourselves to the
BMN operators that correspond to free string states.  In the
closed string sector, an integrable structure is inherited from
its parent theory, $\CN{4}$ SYM. For the open string sector, the
planar one-loop mixing matrix for gauge invariant holomorphic
scalar operators is identified with the Hamiltonian of an
integrable $SU(3)$ open spin chain. Using the $K$-matrix formalism
we identify the integrable open-chain boundary conditions that
correspond to string boundary conditions. The solutions to the
algebraic Bethe ansatz equations (ABAE) with a few impurities are
shown to recover the anomalous dimensions that exactly match the
spectrum of free open string in the plane-wave background. We also
discuss the properties of the solutions of ABAE beyond the BMN
regime.}

 \preprint{USTC-ICTS-04-01}

\begin{document}

\section{Introduction}

Starting from BMN's proposal \cite{BMN02} on duality between the
string theory on a plane-wave background and Super Yang-Mills
(SYM), a number of aspects of this duality have been thoroughly
investigated (e.g., see a recent review \cite{SS03} and references
therein). Compared with the well-known supergravity/CFT approach
\cite{GKP98,Witten98} to AdS/CFT duality \cite{Maldacena1}, the
plane-wave/SYM duality has several remarkable features: 1) In the
plane-wave limit of the $AdS_5\times S^5$ background \cite{BFHP},
the string theory is exactly solvable in light-cone gauge
\cite{Metsaev02,MT02}. This makes it possible to set up the
explicit correspondence between string theory and certain sectors
in large N gauge theory, beyond the gravity/CFT correspondence. 2)
The plane-wave/SYM duality is perturbatively accessible from both
sides. 3) On the SYM side, there exists a double scaling limit, in
which one takes the rank $N$ of the gauge group and the $R$-charge
$J$ of the BMN operators to infinity simultaneously, while keeping
the effective coupling $\lambda'=\lambda/J^2$, rather than the 't
Hooft coupling $\lambda$, fixed. This provide us a chance to study
non-planar contributions in large N Yang-Mills, which correspond
to string interactions in the plane-wave background.

More precisely, the tests of the duality were based on comparing
the spectrum of string excitations with anomalous dimensions of
the corresponding BMN operators. One difficult point in the study
was that there exists operator mixing after taking quantum
corrections into account. To construct generic BMN operators in
SYM and evaluate their anomalous dimensions, one needed to
consider one-loop mixing among a large number of gauge invariant
operators \cite{Beisert02}. A remarkable development in overcoming
this difficulty was the observation made in \cite{MZ02} that the
planar one-loop mixing matrix for anomalous dimensions in the
scalar sector of $\CN{4}$ $U(N)$ SYM,
\beq{1}
\Gamma_c=\frac{\lambda}{16\pi^2}\sum_{l=1}^L
(K_{l,l+1}+2-2P_{l,l+1}),
\eeq
can be identified with the Hamiltonian of an integrable $SO(6)$
spin chain. Here $\lambda=\gym^2 N$ is the 't Hooft coupling, $K$
and $P$ are the trace and permutation operator, respectively. This
Hamiltonian acts on the Hilbert space ${\cal H}=\otimes_{l=1}^L
{\cal H}_l,\;{\cal H}_l=\mathbb{R}^6$ and satisfies periodic
boundary conditions with ${\cal H}_{L+1}={\cal H}_1$ (forming a
closed chain). Then a powerful tool, the Bethe Ansatz \cite{ABAE}
for one-dimensional (quantum) integrable systems, can be applied
to find the spectrum of BMN operators and match them with string
predictions. In addition, the algebraic Bethe ansatz equation
(ABAE) can be solved even in the case with finite $J$. (For
further developments using the dilatation operator of SYM, see
ref. \cite{Beisert03}.)

More astoundingly, the integrable structure (1) in $\CN{4}$ SYM
has far-reaching implications for strings beyond the plane-wave
limit. Alternatively, the BMN states could be viewed \cite{GKP02}
as quadratic fluctuations of a semi-classical string in
$AdS^5\times S^5$. In the BMN case, the semi-classical solution is
near-BPS. However, the BPS conditions turned out to be not
essential, and there exist semi-classical sectors, far from BPS,
which could also be used for precise tests of AdS/CFT
correspondence. The duality between these so-called spinning
strings and ``long'' scalar composite operators in gauge theory
has been confirmed with the help of solving the ABAE of $SO(6)$
spin chain in the thermodynamic limit. (For a recent review, see
\cite{Tseytlin} and references therein). Moreover, the integrable
spin chains in Wess-Zumino models and orbifold gauge theories with
less supersymmetries ($\CN{1,2}$), even away from the conformal
points, have been studied by two of us in ref. \cite{WW03}.

On the other hand, it is very appealing to add D-branes and study
the dynamics of open strings in the $AdS_5\times S^5$ plus D-brane
backgrounds. This effectively corresponds to adding fundamental
flavors to the dual gauge theory. Then issues similar to the
original BMN proposal for closed strings can be investigated for
open strings too. A well-known model has been proposed in
\cite{BGMNN02}. The gauge theory is a four dimensional $\CN{2}$
$Sp(N)$ gauge theory with matter in $[2]\oplus (4\times 2{\bf N})$
representations. The dual strings are in the plane wave limit of
$AdS_5 \times S^5/Z_2$, arising as the near horizon limit of
D3-branes at an O7-plane in type I' string theory. Various aspects
of the plane-wave/SYM duality in this system have been
investigated: the free plane-wave string and the corresponding
BMN-like operators \cite{BGMNN02, Imamura}, the plane-wave string
interactions in terms of light-cone open-closed string field
theory \cite{BN02,CK03,Stef03,GMP03}, and the semi-classical open
string solutions in $AdS_5 \times S^5/Z_2$ as well \cite{Stef03b}.
(Other studies on D-branes in the plane-wave background can be
found in \cite{Green,LP02,ST02,BHLN02,NSW03,SY03}.)

In this paper, we would like to explore the power of the
integrable structures on the dual $Sp(N)$ gauge theory side. As a
first step in this direction, we will concentrate on the test of
(free) string spectrum, with the contributions corresponding to
string interactions turned off. Then the sectors corresponding to
closed string and open string can be treated independently. The
closed string scalar sector in this theory has the same integrable
structure as in the parent theory $\CN{4}$ SYM; i.e. the planar
one-loop anomalous dimension matrix (ADM) can be identified as the
Hamiltonian of a closed $SO(6)$ spin chain. As for the open string
sector, we restrict ourselves to the operators consisting of
holomorphic scalars except the ``quarks" at the two
ends\footnote{For early study of higher-twist operators in QCD in
terms of integrable open spin chains, see e.g. \cite{Belitsky}.}.
The ADM for such operators is of the form of the Hamiltonian of an
open $SU(3)$ spin chain with boundary terms\footnote{In accordance
with the conventions in the literature of the integrable systems,
we use the symmetry of the bulk part of the spin Hamiltonian (i.e.
the group associated with the Yang-Baxter $R$-matrix) to label the
spin chain, which is $SO(6)$ and $SU(3)$ respectively for the
closed and open chain, though the $Sp(N)$ gauge theory at hand
does not possess these (global) symmetries.}. We have been able to
identify the boundary conditions that are appropriate for those of
open strings and show that they indeed make the open spin chain
integrable. Then the open spin chain can be solved with the help
of the Bethe ansatz. The solutions to the ABAE with a few
impurities will be shown to exactly match the spectrum of a free
open plane-wave string. These results are viewed as a preliminary
step towards studying open string interactions in an approach that
exploits integrable structures.

Though we will work in the context of the $Sp(N)$ gauge theory,
the main techniques and results are expected to be applicable to
other $\CN{2}$ SQCD \cite{Imamura}.

The paper is organized as follows: In Section 2, we first give a
short review of the $\CN{2}$ $Sp(N)$ theory and its duality to
plane-wave strings; then we study planar one-loop operator mixing.
In Section 3, we proceed to uncover the integrable structures in
this model, paying particular attention to the sector dual to free
open string. Then we solve the resulting integrable open spin
chain with the Bethe ansatz, and show that the solutions to the
ABAE agree with string theory predictions. Finally we conclude the
paper with discussions in Section 4.

\section{Operator Mixing in $\CN{2}$ $Sp(N)$ theory}

The $\CN{2}$ $Sp(N)$ theory we study in this paper arises from the
orientifold projection of 2N D3-branes spreading along
$x_1,x_2,x_3$ directions, in the presence of four D7-branes and an
O7-plane sitting at the origin $x_7=x_8=0$. Due to the orientifold
action, the strings become unoriented, and the gauge group on
D3-branes changes from U(2N) to $Sp(N)$. The $Sp(N)$ gauge theory
is superconformal when all 7-branes and D3-branes are sitting
together. In this case, the near horizon geometry of D3-branes is
$AdS_5 \times S^5/Z_2$. One then takes the plane-wave limit by
boosting along a circle inside $S^5/Z_2$ in a world-volume
direction of the D7-branes. The D7-branes together with 5-form
flux break $SO(8)$ symmetry down to $SO(4)\times SO(2)\times
SO(2)$, where two $SO(2)$ act on $z'=x_5+ix_6$ plane and
$w=x_7+ix_8$ plane respectively. Therefore, the Neumann directions
are along $\{x^1,\cdots,x^4,z',\bar{z}'\}$ and the Dirichlet
directions are along $\{w,\bar{w}\}$. The light-cone spectrum of
the free string in the plane wave limit of $AdS_5 \times S^5/Z_2$
is given by \cite{BGMNN02,Imamura}
\beq{2}\frac{2}{\mu}p^-=\Delta-J=\left\{\begin{array}{ll}
\displaystyle \sum_{n=-\infty}^{+\infty} N_n\sqrt{1+\frac{4\pi
g_sNn^2}{J^2}}\quad\quad\quad\quad\quad\quad
&{\rm for\;\;closed\;\;string,}\\
\displaystyle 1+\sum_{n=-\infty}^{+\infty} N_n\sqrt{1+\frac{\pi
g_sNn^2}{J^2}}&{\rm for\;\;open\;\;string.}\end{array}\right.
\eeq

The dual $D=4,\;\CN{2}$ $Sp(N)$ gauge theory has $SU(2)_R\times
U(1)_R$ R-symmetry and $SU(2)_L\times SO(8)$ global symmetry. The
bosonic field content consists of the following $\CN{2}$
supermultiplets: the vector multiplet $(V,W)$ in the adjoint
representation, a hypermultiplet $(Z,Z')$ in the antisymmetric
representation, and four hypermutiplets $(\td{q}_A,q_A)$ in the
fundamental representation ($A=1,...,4$). In $\CN{1}$ language,
$V$ is a vector multiplet, while $W,\;Z,\;Z',\;\td{q}_A,\;q_A$ are
all chiral multiplets. The superpotential reads
\beq{3} {\cal W}\sim q_AW\td{q}_A+{\rm
tr}[(\Omega W)(\Omega Z)(\Omega Z')],
\eeq
where $\Omega$ is the invariant rank-2 tensor of $Sp(N)$.

Now let us turn to the mixing matrix for anomalous dimension of
composite operators in this $\CN{2}$ $Sp(N)$ gauge theory. As in
refs. \cite{MZ02,WW03}, we consider gauge invariant composite
operators consisting of various chiral scalar matter fields
without derivatives. The operator basis corresponding to closed
string states are the single trace operators
\beq{4} {\cal O}_{i_1,...,i_{_L}}^{close}={\rm tr}[
(\Omega\Phi_{i_1})\cdots(\Omega\Phi_{i_{_L}})],
\eeq
while the operator basis corresponding to open string states is of
the form
\beq{5} {\cal O}_{i_1,...,i_{_L}}^{open}=\lambda_{pq}Q^q\Omega
(\Phi_{i_1}\Omega)\cdots(\Phi_{i_{_L}}\Omega)Q^p.
\eeq
Here the $Q^p\;(p=1,...,8)$ are linear combinations of the
``quarks'' $(\td{q}_A,q_A)$ forming an $SO(8)$ vector,
$\lambda_{pq}$ the Chan-Paton factors, and $\Phi_i$ linear
combinations of $Z,\;Z'$, $W$ and their complex conjugates forming
an ``$SO(6)$ vector''\footnote{Strictly speaking, their $SO(6)$
rotations do not form a global symmetry in this theory, because
$W$ and $(Z,Z')$ belong to different representations of gauge
group $Sp(N)$. However, in the planar diagrams, they give rise to
symmetric contributions to the operator mixing matrix (or ADM),
because of a formal SO(6) invariance for the second term of the
superpotential (\ref{3}). Therefore an $SO(6)$ symmetry will
appear in the spin chain Hamiltonian that corresponds to the
planar one-loop ADM. The fact that the symmetry of the integrable
spin chain is not directly related to the global symmetry of the
theory, rather it is determined by the formal invariance of the
relevant superpotential, has been already observed in ref.
\cite{WW03}.}. In addition to having the ``quarks'' $Q^p$ at the
ends in operator basis~(\ref{5}), ``anti-quarks'' $\bar{Q}^p$ are
also allowed to be put at the ends. At the leading order in $1/L$,
the operator basis with $Q^p$ at ends and that with $\bar{Q}^p$ at
ends do not mix\footnote{The mixing between the two operator bases
may occur only when a $\bar{Z}'$ impurity appears in the
neighborhood of boundary (anti-)quarks and, hence, is expected to
be sub-leading in $1/L$\cite{Imamura,GMP03}.}. So we can treat
them separately. For simplicity, we will consider only the
operator basis~(\ref{5}) with $Q^p$ at the ends.

Concerning the closed string sector, two facts are notable: 1)
Chiral matters $W,Z,Z'$ are obtained by a $Z_2$ orientifold
projection from scalar fields in the $\CN{4}$ $U(2N)$ theory. 2)
All correlation functions of the orbifolded theory are known to
coincide with those of its parent $\CN{4}$ SYM \cite{BZV98,BJ98},
except for a combinatoric factor. Consequently the closed-string
BMN operators, i.e. linear combinations in the operator
basis~(\ref{4}), can be obtained by a $Z_2$ orientifold projection
from those in $\CN{4}$ $U(2N)$ SYM. The planar one-loop anomalous
dimension matrix (ADM) for these operators is expected to coincide
with matrix~(\ref{1}) by replacing $\lambda\to 2\lambda$
\cite{MZ02,WW03}. However, this statement is not exactly true for
our $Sp(N)$ theory, because there is mixing between operators in
${\cal O}^{close}$ and in ${\cal O}^{open}$. For example, the
two-point function of the operators ${\cal O}_1\;\sim\; {\rm
tr}[(\Omega W)(\Omega W^{\dag})(\Omega Z)^J]$ and ${\cal O}_2
\;\sim\;Q^q\Omega(Z\Omega)^JQ^q$ is non-zero even at the planar
one-loop level. In fact, it is of the order ${\cal O}(\sqrt{g_2})$
with $g_2=J^2/N$, characteristic of open string
interactions\footnote{The details for power counting can be found
in ref.\cite{GMP03}. It should be pointed out that this ${\cal
O}(\sqrt{g_2})$ counting is not originated from non-planar
diagrams, but from the difference in normalization between closed-
and open-string operators: We normalize ${\cal O}_1$ by
multiplying $\gym^{-(L+2)}N^{-(L+2)/2}$ and ${\cal O}_2$ by
multiplying $\gym^{-(L+2)}N^{-(L+1)/2}$. Such normalization makes
the two-point functions $\langle{\cal O}_1{\cal\bar O}_1\rangle$
and $\langle{\cal O}_2{\cal\bar O}_2\rangle$ both independent of
the 't Hooft coupling at tree level. Then, the usual 't Hooft
counting shows that $\langle{\cal O}_1\bar{\cal
O}_2\rangle\;\sim\; \lambda/\sqrt{N}$.}. It reflects the fact that
7-7 strings interact with closed strings in the bulk. Since our
goal in this paper is to explore the power of integrable
structures in studying free string spectrum, these contributions
in gauge theory representing string interactions are not relevant.
So we temporarily turn them off, and hope to incorporate them back
when we study string interactions in the future.

Once turning off the contributions dual to open-closed string
interactions, the closed and open string sectors (\ref{4}) and
(\ref{5}) are decoupled at the planar one-loop level, and we are
allowed to treat the free closed string and free open string
separately. Then the discussion for the closed string scalar
sector is the same as in \cite{MZ02}, so one has an $SO(6)$ closed
spin chain, whose Hamiltonian corresponds to the ADM of
single-trace BMN operators.

Let us turn to the open string sector. The key difference between
open and closed string BMN operators is that the latter consists
of single trace operators, while the former operators with
fundamental ``quarks'' appearing at the two ends. So we naturally
expects that open-string BMN operators correspond to an open spin
chain with boundary. For simplicity, we will focus on the gauge
invariant operators consisting of holomorphic scalars. Recall that
the $\CN{4}$ theory can be represented in terms of $\CN{1}$ fields
with manifest $SU(3)$-invariant superpotential ${\cal W}\sim {\rm
tr}(Z_1[Z_2,Z_3])$, where $Z_i=\Phi_{2i-1}+i\Phi_{2i}$, and
$\Phi_i$ form an $SO(6)$ vector. Then the gauge invariant
composite operators consisting of only $Z_i$ (or of only
$Z_i^\dag$) form the (anti-)holomorphic class, and the rest
containing both $Z_i$ and $Z_i^\dag$ form the non-holomorphic
class. The crucial point is that operators belonging to different
classes do not mix each other under planar one-loop corrections.
It has been shown \cite{WW03} that the Hamiltonian restricted to
the holomorphic class,
\beq{6} \Gamma_{c}=\frac{\lambda}{4\pi^2}
\sum_{l=1}^L (1-P_{l,l+1}),
\eeq
describes an integrable spin chain with $SU(3)$ symmetry. One can
do the similar classification in the $\CN{2}$ $Sp(N)$ theory, for
both closed string and open string sectors. Up to planar one-loop,
the ADM for the open-string operators~(\ref{5}) in the holomorphic
class can be easily computed using Feymann diagrams (Fig. 1), with
the result
\beq{7}
\Gamma_o=\frac{\lambda}{4\pi^2}\sum_{l=1}^{L-1} (1-
P_{l,l+1})+\frac{\lambda}{4\pi^2} (\Sigma_1+\Sigma_L).
\eeq
Here the boundary terms are $\Sigma_1=\Sigma(\otimes I_{3\times 3}
)^{L-1},\;\Sigma_L=(I_{3\times 3}\otimes)^{L-1}\Sigma$, with
$\Sigma={\rm diag}\{0,0,1\}$. The ``bulk'' part (the first term)
in $\Gamma_o$ is the same as in the closed string sector. It is
from three sources: flavor-blind gauge boson propagation, hopping
of the impurity induced by the interactions between the chiral
fields and self-energy corrections of the chiral fields. (See
\cite{WW03} for details.) The last term in $\Gamma_o$ exhibits the
effects of the ``quarks'' at the ends. The nonzero element in
$\Sigma$ is a manifestation that only when the chiral field at the
first or the last site is $W$, the ADM receives an extra
contribution from the first term in the superpotential~(\ref{3})
(see fig. 1). This corresponds to the Dirichlet boundary condition
for open string \cite{BGMNN02}. And the zero diagonal components
in $\Sigma$ correspond to the Neumann boundary condition.

\begin{figure}[hptb]
\centering
\includegraphics[width=5.5in]{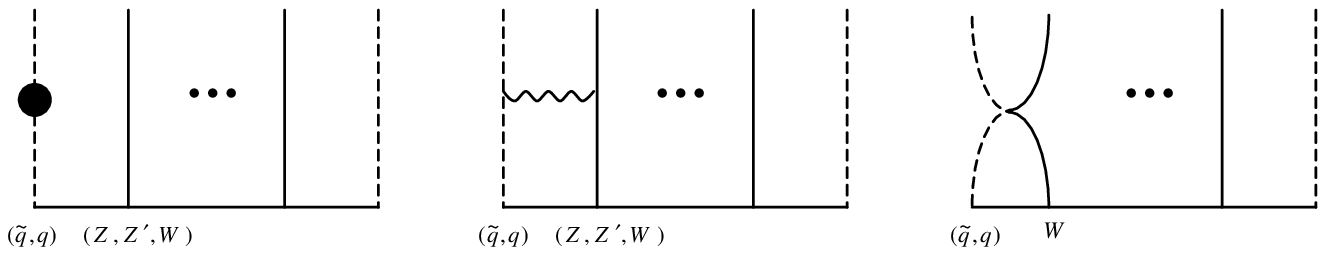}
\begin{minipage}{5in}
\caption{One-loop planar diagrams involving quarks at boundary.
Here the solid lines denote chiral fields $(Z,Z',W)$ and dash
lines denote the chiral field $(\td{q},q)$. It should be noticed
that $Z,\;Z'$ contribution is absent in the third graph.}
\end{minipage}
\end{figure}

The ADM $\Gamma_o$ is of the form of the Hamiltonian of an open
$SU(3)$ spin chain\footnote{Note that the $SU(3)$ symmetry of the
``bulk'' part of the open spin chain is not a symmetry of the
gauge theory, for the same reason as pointed out in footnote 3.
Moreover, as emphasized in footnote 4, this result holds only when
we ignore the mixing with BMN operators with ``anti-quarks'' at
the ends, which is of order of $1/L$.}. In next section we will
prove that the boundary terms make the open spin chain integrable,
and try to solve it by using the Bethe ansatz, showing that the
solutions agree with the spectrum of the plane-wave open string.

\section{Integrable Open Spin Chain and Its Solutions}

Recall that the (quantum) integrability of closed spin chains is
guaranteed, if there is an $R$-matrix satisfying the Yang-Baxter
equations \cite{Yang67,Baxt70}
\beq{8}
R_{ab}(u)R_{ac}(u+v)R_{bc}(v)=R_{bc}(v)R_{ac}(u+v)R_{ab}(u),
\eeq
where $R_{ab}$ acts on the tensor space ${\cal H}_a\otimes {\cal
H}_b$, with ${\cal H}_a$ the Hilbert space associated with the
$a$-th lattice site, and $u$ or $v$ the spectral parameter. The
explicit expression of the $R$-matrix depends on the symmetry
group and the representation of the spin chain. For the $SU(n)$
chain (in the fundamental representation), it reads
\cite{Yang67,Suth70}
\beq{9}R_{ab}(u)=\frac{1}{u+i}[uI_{ab}+iP_{ab}]. \eeq

For open spin chains, the integrability requires, in addition to
an $R$-matrix satisfying eq.~(\ref{8}), also the existence of a
boundary $K$-matrix satisfying the boundary Yang-Baxter equation
\cite{Sklyanin88,BYB}
\beq{10}R_{ab}(u-v)K_a(u)R_{ba}(u+v)K_b(v)=K_b(v)R_{ab}(u+v)K_a(u)R_{ba}(u-v).
\eeq

For an SU(3) open chain, the general diagonal $K$-matrix that
solves Eq.~(\ref{10}) (with the $R$-matrix~(\ref{9})) has been
obtained in \cite{VG94}:
\beq{11}K^{\pm}_{(l)}(u,\xi_{\pm})={\rm
diag} \{\overbrace{a^\pm,...,a^\pm}^{l},
\overbrace{b^\pm,...,b^\pm}^{n-l}\}, \eeq where
\beq{12}a^+&=&i(\xi_+-n)-u,\hspace{0.55in}b^+=i\xi_++u,
\nonumber\\
a^-&=&i\xi_-+u,\hspace{1in}b^-=i\xi_--u,
\eeq
with arbitrary $\xi_\pm$ and any $l\in\{1,...,n-1\}$. Here we have
used $\pm$ to label the two boundaries. The corresponding
integrable Hamiltonian is given by \cite{Sklyanin88,DN98}
\beq{13}H_{open}=\sum_{m=1}^{L-1}H_{m,m+1}
+\frac{1}{2\xi_-}\left.\frac{d}{du}
K^-_{1,(l)}(u,\xi_-)\right|_{u=0} +\frac{{\rm
tr}_0[K^+_{0,(l)}(0,\xi_+)H_{L,0}]} {{\rm tr}K^+_{(l)}(0,\xi_+)},
\eeq
where $H_{m,m+1}=I_{m,m+1}-P_{m,m+1}$. $K_{i,(l)}$ denotes the
$K$-matrix acting on Hilbert space ${\cal H}_i$; namely
$K_{1,(l)}$ acts on the Hilbert space at the first site while
$K_{0,(l)}$ on the auxiliary Hilbert space. The eigen-energies and
momenta (which now label standing, rather than traveling, wave
modes), defined relative to the pseudo-vacuum
$\omega=(\vec{v}\otimes)^{L-1}\vec{v}$ (with
$\vec{v}=(1,0,...,0)^T$), are given by \cite{VG94}
\beq{14}
E&=&\sum_{j=1}^{n_1}\ep(\mu_{1,j})+\ep_0(\xi_+,\xi_-),
\hspace{0.5in} \ep(\mu)=\frac{4}{\mu^2+1},\nonumber \\
P&=&\sum_{j=1}^{n_1}p(\mu_{1,j}), \hspace{1.3in}
p(\mu)=\frac{1}{i}\ln\frac{\mu+i}{\mu-i}.
\eeq
Here $\mu_{i,j}$ satisfy the algebraic Bethe ansatz equations
(ABAE)
\beq{15}1&=&[e_{l-2\xi_-}(\mu_{l,k})e_{2\xi_+-l}(\mu_{l,k})\delta_{l,q}
+(1-\delta_{l,q})]\prod_{j=1}^{M_{q-1}}e_{-1}(\mu_{q,k}-\mu_{q-1,j})
e_{-1}(\mu_{q,k}+\mu_{q-1,j}) \nonumber \\
&&\times \prod_{j=1,\atop j\neq k}^{M_q}e_{2}(\mu_{q,k}-\mu_{q,j})
e_{2}(\mu_{q,k}+\mu_{q,j})
\prod_{j=1}^{M_{q+1}}e_{-1}(\mu_{q,k}-\mu_{q+1,j})
e_{-1}(\mu_{q,k}+\mu_{q+1,j}) \\
&&
 \mbox{for $k=1,\cdots, M_q$ and $q=1,\cdots, n-1$.} \nonumber
\eeq
Here $M_0=L,\;M_n=0,\;\mu_{0,j}=\mu_{n,j}=0$, and
\beq{16}
e_n(\mu)=\frac{\mu+in}{\mu-in}.
\eeq

In our present case, $n=3$. We choose $l=2$ (there is a symmetry
between $l=2$ and $l=1$). Then we see that with $\xi_+=\xi_-=1$,
the Hamiltonian~(\ref{13}) reduces to
\beq{17}
H_{open}=\sum_{l=1}^{L-1} (1-P_{l,l+1}) +(\Sigma_1+\Sigma_L)
+\frac{1}{6}.
\eeq
This is nothing but the ADM $\Gamma_o$ given by eq. (\ref{7}), up
to a constant factor! So the open $SU(3)$ spin chain (\ref{7}) is
indeed an integrable system and can be solved exactly.

The parameters $\xi_{\pm}$ specify the boundary conditions in a
general open spin chain model. In the case at hand, it breaks the
bulk $SU(3)$ symmetry down to $SU(2)\times U(1)$, the same as the
$R$-symmetry of the gauge theory at hand. Now the ABAE~(\ref{15})
reduce to
\beq{18}[e_1(\mu_{1,k})]^{2L}&=&\prod_{j=1,\atop j\neq k}^{M_1}
e_2(\mu_{1,k}-\mu_{1,j}) e_{2}(\mu_{1,k}+\mu_{1,j})
\prod_{j=1}^{M_2}e_{-1}(\mu_{1,k}-\mu_{2,j})
e_{-1}(\mu_{1,k}+\mu_{2,j}), \nonumber \\
1&=&\prod_{j=1\atop j\neq k}^{M_2} e_2(\mu_{2,k}-\mu_{2,j})
e_{2}(\mu_{2,k}+\mu_{2,j})
\prod_{j=1}^{M_1}e_{-1}(\mu_{2,k}-\mu_{1,j})
e_{-1}(\mu_{2,k}+\mu_{1,j}).
\eeq

Before we look for the solutions of ABAE, let us clarify a few
points concerning the open spin chain. As we have seen, the
``quarks'' serve as the boundary fields on the chain, while each
chiral scalar field in the composite operators, one of the
$(\Omega Z,\Omega Z',\Omega W)$, corresponds to a site in the spin
chain. The ground state associated with the Hamiltonian~(\ref{7})
is
\beq{19}
G\;\sim\;Q^p\Omega(Z\Omega)^{J-1}Q^q.
\eeq
It corresponds to the open string state with $\Delta-J=1$ and is
an eigenstate of the Hamiltonian~(\ref{7}) with zero eigenvalue.
On the other hand, in the vector notation, it corresponds to the
pseudo-vacuum $\omega$ used to construct the ABAE for the open
spin chain. Therefore, the ground state of Hamiltonian~(\ref{17})
must also have zero energy. This fact, together with
Eq.~(\ref{14}), implies that we should take
\beq{20}
\ep_0(\xi_+=\xi_-=1)=-\frac{1}{6}.
\eeq

For the present spin chain there are two types of impurities,
labeled by two rapidities $\mu_{1,j}$ and $\mu_{2,j}$, which are
associated with the two simple roots $\vec{\alpha}_1$ and
$\vec{\alpha}_2$ of Lie algebra $SU(3)$. The states with
impurities correspond to the excitations of $Z'$ and $W$ above the
ground state $G$, i.e. the BMN operators with the replacements of
some $Z$'s by $Z'$'s and/or by $W$'s in $G$. Consider the highest
weight $\vec{w}$, which generates the fundamental representation
of $SU(3)$. Because the weight $\vec{w}-\vec{\alpha}_2$ is not
equivalent to $\vec{w}$, a single $\mu_2$-impurity without being
bound to a $\mu_1$-impurity on the same site is not allowed. The
physical interpretation is the following: A single
$\mu_1$-impurity ($\vec{w}-\vec{\alpha}_1$) creates a $Z'$
replacement in the state $G$, while a $\mu_1-\mu_2$ bound impurity
($\vec{w}-\vec{\alpha}_1-\vec{\alpha}_2$) creates a $W$
replacement. But an individual $\mu_2$-impurity
($\vec{w}-\vec{\alpha}_2$) would kill the vacuum.

In contrast to closed spin chains, the so-called trace condition,
which expresses the cyclic symmetry of closed-string BMN
operators~(\ref{4}) and reflects the level matching condition of
closed string, is absent for open chains. Hence the first
non-trivial case for the open spin chain is a single
$\mu_1$-impurity in the ground state. It describes the BMN
operators with a single $Z'$-insertion with $\Delta-J= 2$ (or
$L-J=1$ with $L$ the total number of fields $Z,\;Z'$ and $W$), or
an oscillator excitation of the open string in the Neumann
directions. The ABAE now reduces to
\beq{21}\left(\frac{\mu_1+i}{\mu_1-i}\right)^{2L}=1.
\eeq
The solution is $\mu_1=\cot{n\pi/2L}$ which, together with
Eqs.~(\ref{7}), (\ref{14}) and (\ref{17}), gives the anomalous
dimensions
\beq{22}
\gamma_{_{Z'}}=\frac{\lambda}{\pi^2}\sin^2{\frac{n\pi}{2L}} \quad
\xrightarrow{n\ll L=J+1}\quad \frac{n^2\lambda}{4J^2}=\frac{\pi
g_sN n^2}{2J^2}.
\eeq
This result precisely agrees with the open string spectrum with
one oscillator mode when $g_sN n^2/J^2\ll 1$.

The BMN operators with a single $W$-insertion, or the open string
state with a single oscillator mode in the Dirichlet directions,
correspond to a $\mu_1-\mu_2$ bound impurity in the open spin
chain. The ABAE now reduce to
\beq{23}\left(\frac{\mu_1+i}{\mu_1-i}\right)^{2L}&=&
\frac{\mu_1-\mu_2-i}{\mu_1-\mu_2+i}\frac{\mu_1+\mu_2-i}
{\mu_1+\mu_2+i}, \nonumber \\
1&=&\frac{\mu_2-\mu_1-i}{\mu_2-\mu_1+i}\frac{\mu_2+\mu_1-i}
{\mu_2+\mu_1+i}. \eeq The solution
\beq{24}
\mu_1=\cot{\frac{n\pi}{2(L+1)}},\hspace{1in} \mu_2=0
\eeq
yields the anomalous dimensions
\beq{25}
\gamma_{_{W}}=\frac{\lambda}{\pi^2}\sin^2{\frac{n\pi}{2(L+1)}}
\quad \xrightarrow{n\ll L=J+1}\quad \frac{\pi g_sN n^2}{2J^2}.
\eeq
It is remarkable that the anomalous dimensions~(\ref{25}) coincide
with those in Eq. (\ref{22}) for very large $L$ (or $J$), exactly
as anticipated by the open string spectrum~(\ref{2}).

If all of the rapidities $\mu_{q,j}$ are real, it is convenient to
take the logarithm of the ABAE~(\ref{18}): \beq{26}
2L\vartheta(\mu_{1,j})=Q_{1,j}\pi &+&\sum_{k\neq
j}^{n_1}[\vartheta(\frac{\mu_{1,j}-\mu_{1,k}}{2})
+\vartheta(\frac{\mu_{1,j}+\mu_{1,k}}{2})] \nonumber \\
&-&\sum_{k=1}^{n_2}
[\vartheta(\mu_{1,j}-\mu_{2,k})+\vartheta(\mu_{1,j}+\mu_{2,k})],
\nonumber
\\ 0=Q_{2,j}\pi &+&\sum_{k\neq j}^{n_2}[
\vartheta(\frac{\mu_{2,j}-\mu_{2,k}}{2})
+\vartheta(\frac{\mu_{2,j}+\mu_{2,k}}{2})] \nonumber \\
&-&\sum_{k=1}^{n_1}[\vartheta(\mu_{2,j}-\mu_{1,k})
+\vartheta(\mu_{2,j}+\mu_{1,k})],
\eeq
where $\vartheta(x)=\cot^{-1}{(x)}\in (-\pi/2,\pi/2)$, and
$Q_{q,j}$ are integers. Moreover there are no coinciding $Q_{q,j}$
for a given $q=1$ or $q=2$, because the discreteness of the Bethe
roots requires them to be pushed to distinct branches of the
logarithm function.

The above equations in general cannot be solved exactly. In the
thermodynamic limit $L\to\infty$, however, some analytical results
can be worked out. An interesting case is to consider a finite
number of impurities, i.e., $n_1,n_2\ll L$. If $Q_{q,j}/L\ll 1$,
we have approximately $\vartheta(\mu_{1,j})=Q_{1,j}\pi/2L$.
Consequently the anomalous dimensions are approximately given by
\beq{27}
\gamma=\frac{\lambda}{\pi^2}\sum_{j=1}^{n_1} \sin^2{\frac{\pi
Q_{1,j}}{2L}}\quad \xrightarrow{Q_{1,j}\ll L=J+n_1} \quad
\frac{\pi g_sN}{2J^2}\sum_j Q_{1,j}^2.
\eeq
This result matches the free open string spectrum~(\ref{2}) for
states with at most single occupancy of each excited oscillator
mode when $g_sN/J^2\ll 1$.

In contrast to the case with a finite number of impurities, we may
also consider the operator with the largest number of impurities,
i.e., $n_1,n_2\;\sim\; L$. It corresponds to the highest excited
eigenstate of the Hamiltonian~(\ref{7}), with the largest
eigenvalue (anomalous dimension). In this case, we can replace
$Q_{q,j}$ in the ABAE~(\ref{26}) by $j$, and $j/L$ by a continuous
variable $x$. Adopting the procedure presented in
\cite{YY69,MZ02,WW03}, we conclude that there are $2L/3$
$\mu_1$-impurities and $L/3$ $\mu_2$-impurities for the highest
excited state. It implies that there are an equal number of
$Z,\;Z'$ and $W$ fields in the corresponding composite operator,
and they form an $SU(3)$ singlet. Its anomalous dimension is
\cite{WW03}
\beq{28}
\gamma =-\frac{\lambda}{12\pi^2}L(\frac{\pi}{\sqrt{3}}+3\ln{3}).
\eeq
This operator is far from BPS, and is beyond the BMN regime.
Moreover, this one-loop result~(\ref{28}) makes sense even in the
regime with $\lambda L\ll 1$. It corresponds to $R^4\sim g_sN\to
0$ in $AdS_5\times S^5$, just the opposite to the plane wave limit
of $AdS_5\times S^5$.

We also note that the ABAE~(\ref{18}) for the open spin chain
allow complex solutions. As an example, let us consider the case
with two $\mu_1$-impurities. The ABAE now reduce to
\beq{29}
\left(\frac{\lambda_1+i}{\lambda_1-i}\right)^{2L}&=&
\frac{\lambda_1-\lambda_2+2i}{\lambda_1-\lambda_2-2i}
\frac{\lambda_1+\lambda_2+2i}{\lambda_1+\lambda_2-2i},
\nonumber \\
\left(\frac{\lambda_2+i}{\lambda_2-i}\right)^{2L}&=&
\frac{\lambda_1-\lambda_2-2i}{\lambda_1-\lambda_2+2i}
\frac{\lambda_1+\lambda_2+2i}{\lambda_1+\lambda_2-2i},
\eeq
where $\lambda_1=\mu_{1,1},\;\lambda_2=\mu_{1,2}$. When
$\lambda_1$ acquires an imaginary part, the left hand side of the
first equation in (\ref{29}) grows or decreases exponentially as
$L\to\infty$. Therefore, we make the ansatz
\beq{30}
\lambda_1=a+iu,\hspace{1in} \lambda_2=a-iu,
\eeq
where $a$ and $u$ are real and $a\;\sim\; L,\;u\;\sim\; L^k,\;
(k<1)$. The assumption $\lambda_2=\bar{\lambda}_1$ ensures the
energy to be real. Substituting Eq.~(\ref{30}) into the
ABAE~(\ref{29}) and expanding it to the sub-leading order in
$1/L$, we obtain
\beq{36}
\left(1+\frac{2(u-1)}{a^2}+\frac{2i}{a}\right)^{2L}\simeq
\frac{u+1}{u-1}\frac{a+i}{a-i}, \nonumber \\
\left(1-\frac{2(u+1)}{a^2}+\frac{2i}{a}\right)^{2L}\simeq
\frac{u-1}{u+1}\frac{a+i}{a-i}.
\eeq
It can be reduced to
\beq{37}
&&\left(1+\frac{4u}{a^2}\right)^{L}\simeq \frac{u+1}{u-1},
\nonumber \\
&&2L\vartheta(\frac{a}{2})=\left\{\begin{array}{ll} \displaystyle
2m\pi+2\vartheta(a)+O(1/L)\quad\quad\quad\quad
&k>0\\
\displaystyle (2m+1)\pi+2\vartheta(a)+O(1/L)\quad\quad\quad\quad
&k<0 \end{array}\right.
\eeq

For finite $m$, i.e., $m/L\ll 1$, we have $a\sim L$. Then the
first equation in (\ref{37}) gives $k=1/2$. This result is similar
to the closed chain case, in that Bethe roots behave as
$\lambda=c_0L+ic_1\sqrt{L}+...$ in a large $L$ expansion when they
pick up an imaginary part. Finally we get the solution (with $m$
an integer)
\beq{38}
a\simeq 2\cot{\frac{m\pi}{L}}.
\eeq
The energy (or anomalous dimension) of the solution~(\ref{38}) is
\beq{39}
\gamma_{b}=\frac{\lambda}{2\pi^2}\sin^2{\frac{m\pi}{L}} \quad
\xrightarrow{m\ll L=J+2}\quad \frac{\pi g_sN m^2}{J^2},
\eeq
with the right hand side corresponding to open string energy
(\ref{2}) with two excitations of mode number $m$ along the
Neumann direction $z'$. We expect that similar results should be
true for complex roots corresponding to a finite number of
oscillator excitations as $L\to\infty$.

\section{Conclusions and Discussions}

To conclude, in this paper we have investigated the integrable
structures in an $\CN{2}$ superconformal $Sp(N)$ gauge theory in
four dimensions, which is known to be dual to an open+closed
string system. We showed that the planar one-loop ADM associated
with the gauge invariant composite operators~(\ref{4}) or
(\ref{5}) restricted to the holomorphic scalar sector can be
identified with the Hamiltonian of either a closed or open
integrable $SU(3)$ spin chain. In particular, we have established
that the boundary terms in the open string chain in gauge theory
that are appropriate for open string boundary conditions are
indeed integrable ones. The Bethe ansatz method is exploited to
solve the open spin chain. For solutions with a few impurities,
the energy is shown to reproduce the anomalous dimensions of
corresponding BMN operators, in perfect agreement with the free
open string spectrum in a plane-wave background.

Our study raises several interesting questions. As shown in the
last section, there are solutions with a large number of
impurities, which is beyond the BMN regime and far from BPS. From
the lessons in the closed spin chain, one may suspect that such
solution could be related to the AdS/CFT test of certain
semi-classical solutions in string theory; a possible candidate
might be the open spinning strings \cite{Stef03}. A careful
inspection shows that the solution (3.29) does not correspond to
an open spinning string: its energy scales as $L$, while in
general the energy of spinning strings scales as $1/L$. Moreover,
there exist complex roots to the ABAE. We have seen in a simple
example that in the limit $L\to \infty$, they reproduce the free
string spectrum with, say, two excitations of mode number $m$. How
to incorporate the complex roots in the Yang-Yang ansatz
\cite{YY69} for thermodynamics is still an open question. We leave
further study of complex Bethe roots to future research.

In our study, we have been focusing on the gauge invariant
operators in the holomorphic class, which corresponds to the
replacement of $Z'$- and $W$-impurity in a pure $Z$-chain. As we
know, in general $\bar{Z'}$- and $\bar{W}$-impurities are allowed
in gauge theory, breaking the holomorphic nature of the chain.
Such BMN operators have more involved ADM, and gives rise to more
complicated closed or open spin chain. To clarify whether these
spin chains are integrable is also an interesting question.

As mentioned in Section 2, even at the planar one-loop level,
there exist the mixing between closed and open BMN operators. They
naturally correspond to open/closed string interactions in the
system. Though in this paper we have turned off these
contributions in order to explore the power of the integrable
structures in testing the free string spectrum, in particular in
the hope of going beyond the BMN regime. However, the fact that
the planar 1-loop ADM for the gauge invariant composite operators
encodes information on string interactions is quite remarkable.
(Indeed as shown in the light-cone string field theory
\cite{Stef03,GMP03}, open cubic and open-closed string
interactions show up in the planar 1-loop calculation of two-point
functions in SYM.) It would be fascinating to explore effects of
string interactions on the gauge theory side by dealing with a
more complicated or more complete ADM. Also it would be
interesting to see whether there is a bigger integrable structure
in the model at hand, such that the integrable structures we found
in this paper separately for the free closed and open string
sectors are only its substructures.

Of course to pin down the origin of the integrable structure(s),
both on the SYM side and on the dual string theory side, is still
a challenge. It has been revealed in \cite{DNW03,AS03} that there
is a relation between the infinite-dimensional non-local symmetry
of type IIB superstring in $AdS_5\times S^5$ \cite{BPR03,Alday03}
and a non-Abelian and nonlinear infinite-dimensional Yangian
algebra for weakly coupled SCYM. On the other hand, the present
paper provides an example in which the symmetry of the bulk part
of the integrable spin chain corresponding to the planar one-loop
ADM in the BMN scalar sectors, $SO(6)$ (or $SU(3)$) for the closed
(or open) string case, is {\it not} the global symmetry of the
dual $\CN{2},\;Sp(N)$ gauge theory. (Rather, the spin chain
symmetry is originated from a formal invariance of relevant terms
in the superpotential, as observed in ref. \cite{WW03}.) This
seems to be in line with the observation made in \cite{Roiban03},
that the existence of global symmetries in a field theory seems
not an essential ingredient in its relation to an integrable
model. Further study of relevant issues is needed to gain deeper
insights.

\acknowledgments{X.-J.~Wang would like to thank D.-T.~Peng for
discussions on open spin chains. B.~Chen and X.-J.~Wang thank the
organizers of Taiping Lake Workshop on string theory, where this
work was inspired. Y.-S.~Wu thanks the Interdisplinary Center for
Theoretical Study, Chinese Academy of Sciences for warm
hospitality and support during his visit, when the collaboration
began to form. This work is partly supported by the NSF of China,
Grant No. 10305017, and through USTC ICTS by grants from the
Chinese Academy of Science and a grant from NSFC of China. Finally
the authors would like to thank the referee for careful reading
and several comments which are helpful in improving the
manuscript.}


\begin{thebibliography}{99}
\bibitem{BMN02}D. Berenstein, J.M. Maldacena and H. Nastase, {\sl
Strings in flat space and PP-wave from ${\cal N}=4$ super
Yang-Mills}, JHEP {\bf 04} (2002) 013, hep-th/0202021.
\bibitem{SS03}D. Sadri and M.M. Sheikh-Jabbari, {\sl The
plane-wave/super Yang-Mills duality}, hep-th/0310119.\\
J.C. Plefka, {\sl Lectures on the Plane-Wave String/Gauge Theory
Duality}, hep-th/0307101.
\bibitem{GKP98}S.S.\ Gubser, I.R.\ Klebanov and A.M.\ Polyakov,
{\sl Gauge theory correlators from non-critical string theory},
Phys. Lett.\ {\bf B428} (1998) 105.
\bibitem{Witten98}E.\ Witten, {\sl Anti-de Sitter space and
holography}, Adv.\ Theory.\ Math.\ Phys.\ {\bf 2} (1998) 253.
\bibitem{Maldacena1}J.M. Maldacena, Adv. Theor. Math. Phys.
{\bf 2}(1998)231, hep-th/9711200.\\
O. Aharony, S.S. Gubser, J.M. Maldacena, H. Ooguri and Y. Oz,
Phys. Rept. {\bf 323}(2000)183.
\bibitem{BFHP}M. Blau, J. Figueroa-O'Farrill, C. Hull and G.
Papadopoulos, JHEP {\bf 01} (2002) 047, hep-th/0110242; {\sl ibid}
Class. Quant. Grav. {\bf 19} (2002) L87, hep-th/0201081.
\bibitem{Metsaev02}R.R. Metsaev, Nucl. Phys. {\bf B625} (2002) 70,
hep-th/0112044.
\bibitem{MT02}R.R. Metsaev and A.A. Tseytlin, Phys. Rev. {\bf D65}
(2002) 126004, hep-th/0202109.
\bibitem{Beisert02}N. Beisert, C. Kristjansen, J. Plefka,
G.W. Semenoff and M. Staudacher, {\sl BMN correlators and operator
mixing in N=4 Super Yang-Mills theory}, Nucl.\ Phys. {\bf B650}
(2003) 125, hep-th/0208178;\\ N. Beisert, C. Kristjansen, J.
Plefka and M. Staudacher, {\sl BMN gauge theory as a quantum
mechanical system}, Phys.\ Lett. {\bf B558} (2003) 229,
hep-th/0212269.
\bibitem{MZ02}J.A.\ Minahan and K. Zarembo, {\sl The Bethe ansatz
for ${\cal N}=4$ super Yang-Mills}, JHEP {\bf 03} (2003) 013,
hep-th/0212208.
\bibitem{ABAE}H. Bethe, Z. Phys. {\bf 71} (1931) 205; L.D. Faddeev
and L.A. Takhtajan, J. Sov. Math. {\bf 24} (1984) 241; N.Y.
Reshetikhin, Lett. Math. Phys. {\bf 7} (1983) 205; L.D. Faddeev,
hep-th/9605187.
\bibitem{Beisert03}N. Beisert, C. Kristjansen and M. Staudacher,
{\sl The dialtation operator of conformal $\CN{4}$ Super
Yang-Mills Theory}, Nucl. Phys. {\bf B664} (2003) 131,
hep-th/030306;\\ N. Beisert, {\sl The complete one-loop dilatation
operator of N=4 Super Yang-Mills theory}, Nucl.\ Phys. {\bf B676}
(2004) 3, hep-th/0307015; \\ N. Beisert and M. Staudacher, {\sl
The N=4 SYM integrable super spin chain}, Nucl.\ Phys. {\bf B670}
(2003) 439, hep-th/0307042l;\\ N. Beisert, S. Frolov, M.
Staudacher and A.A. Tseytlin, {\sl Precision spectroscopy of
AdS/CFT}, JHEP {\bf 10} (2003) 037.
\bibitem{GKP02}S.S. Guber, I.R. Klebanova and A.M. Polyakov, {\sl
A semi-classical limit of the gauge/string correspondence}, Nucl.
Phys. {\bf B636} (2002) 99, hep-th/0204051.
\bibitem{Tseytlin}A.A. Tseytlin, {\sl Spinning strings and AdS/CFT
duality}, hep-th/0311139.
\bibitem{WW03}X.J.\ Wang and Y.S.\ Wu, {\sl Integrable spin chain
and operator mixing in $\CN{1},2$ supersymmetric theories},
hep-th/0311073.
\bibitem{BGMNN02}D. Berenstei, E. Gava, J. Maldacena, K.S. Narain
and H. Nastase, {\sl Open strings on plane waves and their
Yang-Mills duals}, hep-th/0203249.
\bibitem{Imamura}Y. Imamura, {\sl Open string: BMN operator
correspondence in the weak coupling regime}, Prog.\ Theor.\ Phys.\
{\bf 108} (2003) 1077, hep-th/0208079.
\bibitem{BN02}D. Berenstein and H. Nastase, {\sl On lightcone
string field from super Yang-Mills and holography},
hep-th/0205048.
\bibitem{CK03}B. Chandrasekhar, A.\ Kumar, {\sl D-branes in
PP-wave light cone string field theory}, JHEP {\bf 06} (2003) 001,
hep-th/0303223.
\bibitem{Stef03}B. Stefanski, {\sl Open String Plane-Wave
Light-Cone Superstring Field Theory}, hep-th/0304114.
\bibitem{GMP03}J. Gomis, S. Moriyama and J. Park, {\sl Open+Closed
String Field Theory from Gauge Fields}, hep-th/0305264.\\
J. Lucietti, S. Schafer-Nameki and A. Sinha, {\sl On the exact
open-closed vertex in plane-wave light-cone string field theory},
hep-th/0311231.
\bibitem{Stef03b}B. Stefanski, {\sl Open Spinning Strings},
hep-th/0312091.
\bibitem{Green}O. Bergman, M.R. Gaberdiel and M.B. Green,
JHEP {\bf 0303}(2003)002.
\bibitem{LP02}P.\ Lee and J.W.\ Park, {\sl Open strings in PP-wave
background from defect conformal field theory}, Phys. Rev. {\bf
D67} (2003) 026002, hep-th/0203257.
\bibitem{ST02}K.\ Skenderis and M.\ Taylor, {\sl Branes in AdS and
PP-wave spacetimes}, JHEP {\bf 06} (2002) 025, hep-th/0204054.
\bibitem{BHLN02}V.\ Balasubramanian, M.X.\ Huang and T.S.\ Levi
and A. Naqvi, {\sl Open strings from $\CN{4}$ super Yang-Mills},
JHEP {\bf 08} (2002) 037, hep-th/0204196.
\bibitem{NSW03}S.G.\ Naculich, H.J.\ Schnitzer and N. Wyllard,
{\sl PP-wave limits and orientifolds}, Nucl.\ Phys.\ {\bf B650}
(2003) 43, hep-th/0206094.
\bibitem{SY03}M.\ Sakaguchi, K.\ Yoshida, {\sl D-branes of covariant
AdS superstrings}, hep-th/0310228.
\bibitem{Belitsky}V.M.\ Braun, S.E.\ Derkachov and  A.N. Manashov,
Phys.\ Rev.\ Lett. {bf 81} (1998) 2020-2023; V.M. Braun, S.E.
Derkachov, G.P. Korchemsky and A.N. Manashov, Nucl. Phys. {\bf
B553} (1999) 355; A.V. Belitsky, Phys.\ Lett. {\bf B453} (1999)
59; A.V. Belitsky, Nucl.\ Phys. {\bf B558} (1999) 259; A.V.
Belitsky, Nucl.\ Phys. {\bf B574} (2000) 407; S.E. Derkachov, G.P.
Korchemsky and A.N. Manashov, Nucl. Phys. {\bf B566} (2000) 203.
\bibitem{BZV98}M. Bershadsky, Z. Kakushadze and C. Vafa, {\sl String
expansion as large $N$ expansion of gauge theories}, Nucl.
Phys. {\bf B523} (1998) 59; \\
Z. Kakushadze, {\sl Gauge theories from orientifolds and large $N$
limit}, Nucl. Phys. {\bf B529}(1998) 157.
\bibitem{BJ98}M. Bershadsky, A. Johansen, {\sl Large $N$ limit of
orbifold field theories}, Nucl. Phys. {\bf B536} (1998) 141.
\bibitem{Yang67}C.N.\ Yang, Phys.\ Rev.\ Lett.\ {\bf 19} (1967)
1312.
\bibitem{Baxt70}R.J.\ Baxter, Ann.\ Phys.\ {\bf 70} (1972) 193.
\bibitem{Suth70}B.\ Sutherland, Phys.\ Rev.\ {\bf B12}(1975) 3795.
\bibitem{Sklyanin88}E.K.\ Sklyanin, J. Phys. {\bf A21} (1988)
2375.
\bibitem{BYB}I.V. Cherednik, Theor.\ Math.\ Phys.\ {\bf 61} (1984)
977; L. Mezincescu and R.I.\ Nepomechie, J.\ Phys.\ {\bf A24}
(1991) L17; S. Ghoshal and A.B.\ Zamolodchikov, Int.\ J.\ Mod.\
Phys.\ {\bf A9} (1994) 3841.
\bibitem{VG94}H.J.\ de\ Vega and A.\ Gonz$\acute{\rm a}$lez-Ruiz,
Mod.\ Phys.\ Lett.\ {\bf A9} (1994) 2207.
\bibitem{DN98}A.\ Doikou and R.I.\ Nepomechie, {\sl Bulk and
boundary $S$-matrices for the $SU(N)$ chain}, Nucl.\ Phys.\ {\bf
B521} (1998) 547, hep-th/9813118.
\bibitem{YY69} C.N. Yang nad C.P. Yang, {\sl Thermodynamics of a
One-Dimensional System of Bosons with Repulsive Delta-Functiuon
Interaction}, J. Math. Phys. {\bf 10} (1969) 1115.
\bibitem{DNW03}L. Dolan, C.R. Nappi and E. Witten, {\sl A Relation
between Approachs to Integrabilities in Superconforla Yang-Mills
Theory}, hep-th/0308089.
\bibitem{AS03}G. Arutyunov, M. Staudacher, {\sl Matching Higher
Conserved Charges for Strings and Spins}, hep-th/0310182.
\bibitem{BPR03}I. Bena, J. Polchinski and R. Roiban, {\sl Hidden
symmetries of the $AdS_5\times S^5$ superstring}, hep-th/0305116.
\bibitem{Alday03}L.F.\ Alday, {\sl Non-local charges on
$AdS_5\times S^5$ and PP-waves}, hep-th/0310146.
\bibitem{Roiban03}R. Roiban, {\sl On spin chains and field
theories}, hep-th/0312218.
\end{thebibliography}
\end{document}